\documentclass{aipproc}

\usepackage{graphics}
\usepackage{color}
\usepackage{amsmath}
\usepackage{longtable}

\def\numu{\nu_{\mu}}

\def\nutau{\nu_{\tau}}

\def\inch{$^{\prime\prime}$}

\begin{document}

\title{Feasibility of a Next Generation Underground Water Cherenkov 
Detector: UNO 
\footnote{To be published in the Proceedings of the Next generation
Nucleon decay and Neutrino detector (NNN99) Workshop, Sep. 23-25,
1999, Stony Brook, New York}}

\author{Chang Kee Jung}
\affiliation{The State University of New York at Stony Brook,
Stony Brook, New York 11794-3800, USA}

\begin{abstract} 
The feasibility of a next generation underground water Cherenkov
detector is examined and a conceptual design (UNO) is presented. 
The design has a linear detector configuration with a total volume 
of 650 kton which is 13 times the total volume of the Super-Kamiokande
detector. It  corresponds to a 20 times increase in fiducial volume for
physics analyses. 
The physics goals of UNO are to increase the sensitivity of
the searches for nucleon decays about a factor of ten and to make
precision measurements of the solar and atmospheric neutrino
properties. In addition, the detection sensitivity for Supernova neutrinos 
will reach as far as the Andromeda galaxy.
\end{abstract}

\maketitle
\section{Introduction}

Large scale underground  water Cherenkov detector experiments, 
Super-Kamiokande 
and its predecessors (IMB and Kamiokande),
have been extremely successful in producing crucial physics results
during the last two decades. Their accomplishments includes: the first 
real time measurement of the solar neutrinos, confirmation of the
solar neutrino flux deficit, observation of the neutrino oscillations
in the atmospheric neutrinos, observation of neutrinos from
the Supernova 1987A, and setting the world best limits on the nucleon decays. 

These detectors were originally conceived for  searches for nucleon decays
predicted by various Grand Unification Theories (GUTs). 
While no positive observation of nucleon decays have been made to
date with these detectors, the evidence for neutrino oscillations now
firmly established by the Super-Kamiokande atmospheric neutrino analysis 
provides us with a breakthrough in particle physics beyond the
Standard Model.\cite{nuosc98}
This finding  indicates that the neutrino masses
are indeed very small if we assume no degeneracy in mass eigenstates,
which in turn indicates that there may be a new very high
energy physics scale that facilitates small neutrino masses via 
``See-saw'' mechanism and allows protons to decay.

There are many theoretical 
models that predict proton decays and some of them were 
presented in this workshop (NNN99)\cite{NNN99}.
A specific example of such models can be 
found in Ref.~\cite{Pati} which  presents a complete and 
detailed description of interplay between the neutrino masses,
proton decays and other Standard Model observables in G(2,2,4) and 
SO(10) frameworks. The model predicts proton decay lifetimes within
the reach of the Super-Kamiokande, especially in the SUSY favored 
decay modes. These predictions along with many other predictions from 
other models encourage us to turn our attention back to the
proton decay searches with higher sensitivity.

If discovered, proton decay provides only and unambiguous direct evidence of
existence of GUT scale physics at low energies. It certainly will be
remembered as one of the most revolutionary discoveries in particle
physics history. Some say the discovery is around the corner.

In order to further our effort to search for nucleon decays which I
believe is in the category of ``must-do'' physics and to do high statistics 
studies of neutrino physics including solar neutrinos, atmospheric
neutrinos and supernova neutrinos, I propose a next generation large
underground water Cherenkov detector which is named
UNO (Ultra underground Nucleon decay and neutrino 
Observatory). 

In the following sections, the design
considerations, baseline configuration, physics capabilities and cost
estimation of the UNO detector is described in some detail. 

\section{Ultra underground Nucleon decay and neutrino
Observatory Detector}

The design philosophy of UNO 
is to make a relatively simple extension 
of the well established water Cherenkov detector technology beyond 
Super-Kamiokande to achieve an order of magnitude better sensitivity in
nucleon decay searches and to study various neutrino physics
with higher precision. With water Cherenkov detector technology, we
can utilize the tremendous amount of experience and expertise gained
from the IMB, Kamioka and Super-Kamiokande detectors. 

In order to establish  reasonable detector parameters for physics
capability
studies and cost estimates, I set the benchmark 
 fiducial volume of the UNO detector to be 20 times that of
Super-Kamiokande. The design of the detector 
is kept  to be  simple and robust, 
and  broad physics capabilities are required. 

Several design options are considered keeping in mind 
the practical limitations in conventional water 
Cherenkov detector technique: 
namely, (1) the largest depth of water of a detector
is limited by the  current PMT pressure stress limit (~8 atm for current
20\inch~Hamamatsu PMTs) which can be overcome if one is
willing use an optical high pressure water tight container for each
PMT and compromising the PMT efficiency;
2) The maximum dimension of the detector without 
active detection element is limited by the finite light
attenuation length in pure water ($\sim$80 m @400 nm in Super-Kamiokande ).

Three different detector shapes are considered: Big, Torus and Linear.
According to D. L. Petersen who is a rock engineer at the CNA Consulting
Engineering at Chicago, Illinois,
the excavation costs are more or less same 
independent of the shapes of the detector. \cite{Petersen}
Then, if we want to keep 
the detector cost at minimum 
for a fixed fiducial volume (445 kton; $\sim$20 times the Super-Kamiokande), 
we need to keep the ratio 
$rv$=(fiducial volume/total volume) large and keep
$rs$=(PMT surface area/total volume) small. These requirements  
obviously prefer fat detectors.

The Big detector design option considered 
has a cube shape with 86x86x86m outer dimensions. It has a potential 
problem with the 
PMT pressure limit for the PMTs at the bottom of the detector where
the water pressure will be about 9 atmospheric pressure. It also has a  
diagonal length of 150 m which is much longer than the light attenuation
length currently measured in the pure water. 

The Torus detector design option considered
turns out to be very inefficient in term of the $rv$ value and 
is physically not possible if the cross-section is  
60x60 m for a benchmark detector size.
Even with a 50x50 m cross-section a torus would be too tight, {\it
i.e.} the diameter of the central rock column will be too small to
support the structure. 

Thus, a Torus design will have to have a small cross-section which results in
a small $rv$ and large $rs$ values. For example, a 40x40m cross-section 
torus has $rv$=0.6 compared to 0.7 for a Linear option considered below.

The Linear detector design option considered has a 
60mx60mx180m outer dimensions and it appears to be the most optimal one.
When compartmentalized as three 60mx60mx60m cubes in terms of active 
detection elements, it satisfies all of the requirements mentioned above
and results in a reasonable $rv$ and $rs$ values. The compartmentalization
option naturally makes the detector 
cost more expensive but provides several significant benefits 
compared to an open geometry option. First, it minimizes so-called 
flasher background events, which commonly occur in all water Cherenkov
detectors that use PMTs,
by confining them in an isolated compartment. Second, it
minimizes inefficiency and  variation in  efficiency due to finite 
light attenuation length in water. Third, it keeps 
the detector operational live time to close to 100\%. The current detector
operational live time for Super-Kamiokande is about 90\%. The 10\%
inefficiency is mainly due to detector calibrations. With
compartmentalization we can ensure that at least one compartment will be 
alive during the detector calibration times. Keeping the detector
live time very high (virtually at 100\%) 
is of course very important for physics programs such as a 
supernova watch.

Another issue that needs to be dealt with concerning the 
compartmentalization is whether we should make the divisions among
the compartments rigid so that the water can be filled and
drained independently. Certainly making the divisions rigid walls will make  
any future repairing of the detector easier. 
(It should be noted that it takes two months to drain the current 
Super-Kamiokande tank.) However, it is not clear whether this is
an important enough reason for 
us to consider such an option which will incur substantial additional cost.

One could also argue about whether the next generation detector should 
be built underground or underwater. Indeed there are a few  ideas
presented 
in recent meetings that consider very large underwater detectors.
In my opinion, however, one of the most serious disadvantage of a
underwater detector will be its inaccessibility for calibration and repair.
The experience gained from the Super-Kamiokande tells us that a well
selected and maintained underground
detector environment is a wonderful environment for us to
work and to operate the detector. It provides us an  
easy access to the detector for
various calibrations and for trying out new ideas.
On the other hand, I would assume that precision calibrations of a 
deep underwater detector will be extremely difficult if not
impossible, especially if
one is aiming to do a calibration for solar neutrino measurements
using electron linac or DT generator etc. Thus, for a underwater
detector design to become a viable option, a new sets of elaborate remote
calibration systems has to be developed and tested. In addition, 
location and accessibility of other 
service facilities such as water purification system needs to be
considered, and a practical and realistic solution must be found.
Overall, I expect 
many more technical challenges for a underwater detector than for
a underground detector.

\begin{figure}[t]
\resizebox{\columnwidth}{!}  
{\includegraphics{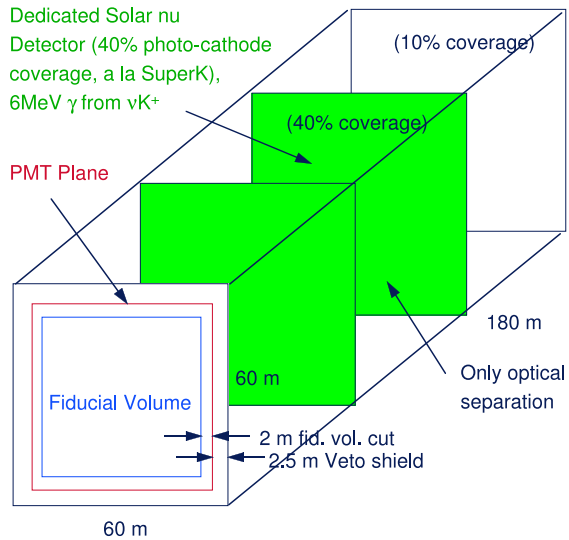}} 
\caption{Baseline configuration of the UNO detector} 
\label{fig:UNO} 
\end{figure}

Considering all of the above issues, I believe that, 
 a large underground 
water Cherenkov detector with a linear configuration is the best
option for a next generation nucleon decay and neutrino detector which
can be built within next 10 years without requiring major R\&D.
A conceptual design of such a detector (UNO detector) is shown in 
Figure~\ref{fig:UNO}. 

The detector has three compartmentalized
sections with 60mx60mx60m dimensions resulting in a total length of
180 m and a total volume of 648 kton. The outer detector region
of the detector has 2.5 m depth of veto shield and is instrumented with 
14,901 8\inch~PMTs with a PMT density of 0.33 PMTs/m$^2$. The inner
detector region has a total fiducial volume of 445 kton which is
defined as the water volume 2 m inside of the inner PMT planes. The
inner detector region is instrumented with 20\inch~PMTs with a
PMT density of 1.96 PMTs/m$^2$ (40\% photocathode coverage a la
Super-Kamiokande) for the central section and 0.49
PMTs/m$^2$ (10\% photocathode coverage) 
for the two sections at the wings. The total number of 
20\inch~PMTs is 56,650 for this configuration. 

\subsection{UNO Physics Goals and Capabilities}

The detector configuration proposed above presents excellent physics
capabilities of UNO for broad ranges of nucleon decay and neutrino 
physics.  The two wings of the detector with 10\% photocathode
coverage will have an energy threshold below 10 MeV which should be
sufficient for nucleon decay searches, supernova neutrino detections,
high statistics atmospheric neutrino studies and high statistics 
solar neutrino studies at the high energy end of
the energy spectrum.   
The central compartment with a 40\% photocathode coverage will serve as  
a dedicated low energy solar neutrino detector with a 5 MeV analysis threshold 
which will also provide
UNO with a detection capability for 
6 MeV $\gamma$'s from  $p\rightarrow\nu K^+$ decays in the oxygen nucleus. 
It also provides the capability to observe 
low energy supernova neutrino events ($\sim 5$ MeV) which are
important for core collapse modeling.  

The primary physics goal of UNO is to obtain 
 an order of magnitude better sensitivity in 
nucleon decay
searches (especially in $p \rightarrow e^+\pi^0$ mode) than Super-Kamiokande.
The exact amount of improvements in the sensitivity vary depending on 
the decay modes, particularly on their expected backgrounds. 
Obviously 
for the background free decay modes we expect more than a order of
magnitude improvements but for the background limited decay modes we
would expect a factor of 4 or 5 improvements. 

The amount of improvement also
strongly depends on the optimization of 
the analyses for high statistics sample. For example, if we continue
current standard Super-Kamiokande analysis for $p \rightarrow e^+\pi^0$, one
will encounter a background limited analysis situation in ten years
(0.5 years for UNO). However, if we make a tighter cut on total
momentum in an event, say 150 MeV (rather than 250 MeV for 
current Super-Kamiokande standard analysis), the signal efficiency
will go down to about 25-30\% from current 44\%, but we will be virtually
background free for many UNO years. Since we gain a factor of 20 in
total fiducial volume, by keeping the efficiency better than 22\% we
can insure a factor of 10 or more gain in the search sensitivity as long
as we can keep the background to be zero. 

In order to estimate the potential improvements reliably, we need to
do much more work. First of all, we need to have better
understanding of the characteristics of the 
atmospheric neutrino background events. This can be
accomplished by studying the K2K 1 kton neutrino events 
which are similar in many aspects to the atmospheric neutrino background events
relevant to nucleon decay searches. Second, we need to reduce the
systematic uncertainties associated with the atmospheric neutrino flux 
and the neutrino interaction cross-section for  water target. Some  
of this information can come from the K2K fine-grained detector data.
Finally, we need to improve the search analyses by developing more
sophisticated algorithms and optimize the analyses for maximum
sensitivity. I personally believe that there is much room for
improvement in these areas, especially for the multi-ring event analysis.
We also need to generate much larger MC events samples with
detailed simulations for these purposes. 
While it is difficult to predict the exact sensitivity of the UNO
detector for nucleon decay searches at this time, 
I am convinced that an order of magnitude improvements in the search
sensitivity for various decay modes are attainable. 

Another equally important physics goal of UNO is 
supernova neutrino detection. With the proposed configuration UNO will 
be able to record 100k neutrino interaction events from a supernova 
explosion at 10 kpc away. This  will allow us to make 
a detailed mapping of the time structure of the neutrino flux providing
valuable information for the theoretical modeling of the supernovae.
The detection reach for neutrinos from supernova explosions
will be about 1 Mpc from the earth which includes most of the local group of
galaxies including Andromeda. We expect (optimistically) 
about one supernova explosion in every ten years within this range.

\begin{figure}[tb]
\resizebox{\columnwidth}{!}{\includegraphics{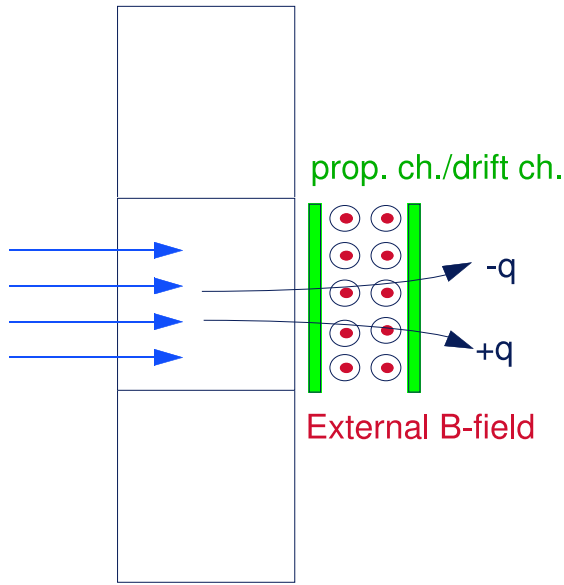}}
\caption{Conceptual layout of UNO as a far detector for a muon
storage ring neutrino beam} 
\label{fig:UNO_far} 
\end{figure} 

Although current Super-Kamiokande measurements of the solar neutrino properties
provide crucial information on solar neutrino problems with
unprecedented statistics, the measurements on  
energy spectrum, day-night flux
ratio, and seasonal variations are limited by the lack of statistics
in providing  an unambiguous resolution to the solar
neutrino problems. Thus, it is strongly desirable to build a large
scale detector that has a capability of precision studies of the
solar neutrinos. 
The proposed configuration of UNO will provide about 20 times more
statistics for high energy end of the solar neutrino events and at
least 7 times more statistics for the low energy end events.
Even if we are able to establish a solution to the solar neutrino
problem with the results from 
Super-Kamiokande, SNO, KAMLAND and Borexino within 
this decade, we will need more precision measurements to firmly
establish the SSM (Standard Solar 
Model) and neutrino oscillation parameters. I believe that
we will be entering a precision measurement era for SSM 
in this decade as the two last decades have been for the
Standard Model.

If we assume the neutrino 
oscillations observed in the Super-Kamiokande atmospheric
neutrino events are due to $\numu$ to $\nutau$ oscillations, there should 
be about 20 tau appearance events produced per year
from the $\nutau$ interactions in the Super-Kamiokande detector. 
For UNO the rate will be about 400 tau appearance events per year.
While extracting a tau signal from background in water Cherenkov
detector is not expected to be easy, it will be a worth while project to
pursue, especially with a reasonably high statistics event sample.
(Currently, we are pursuing such an analysis with the Super-Kamiokande data 
and the initial results
appear to be promising. We are hoping to get about three sigma effect.)
If the $\Delta m^2$ turns out to be  relative low, say 
$\sim 3\times 10^{-3}$ eV$^2$ or lower, the long baseline
neutrino  experiments will have difficult time in observing
unambiguous signal of tau appearance. And thus, establishing a 
direct observation of tau appearance coming from neutrino oscillations
may take much longer time than we currently hope for. 

Finally, a possibility of using UNO  as a far detector of
a Muon Storage Ring produced neutrino beam experiment should be considered.
A water Cherenkov detector is a good candidate for a far detector. It
provides an relatively inexpensive large target volume, particle ID 
for electrons and muons, and reasonable calorimetry. It could also
provide good charge separation if it is aided by an external spectrometer as
shown  in Figure~\ref{fig:UNO_far}.



Thus, if UNO is built, it will serve as a truly multi-purpose general
underground detector for variety of great physics.
Especially for nucleon decay searches, it will serve as a robust
multi-decay mode search detector.

\subsection{UNO Cost Estimates}

In order to obtain a ball park figure cost estimate for UNO
detector, I use the construction cost data for the
Super-Kamiokande detector provided by K. Nakamura as a reference.
I divide the major expense items into two categories according to  
their nature of scaling, i.e. volume-like scaling or surface-like
scaling. And I apply my reasonable guesses to ultimately 
determine the scaling factor from Super-Kamiokande to UNO.  Thus,
these figures should not be taken too
seriously. Table~\ref{tab:cost-estimates}
shows my initial estimate of itemize costs for UNO along with the
actual costs for Super-Kamiokande. The total cost for 
UNO with the baseline configuration proposed in this paper is then
about \$520M and the cost for a detector with 
full 40\% photocathode coverage for all inner surfaces  (a la 
Super-Kamiokande) is about \$680M. And all of these figures do not
include contingencies.

\begin{table}[htbp] 
    \begin{tabular}{lrcr}
\hline\hline 
      Item                     & SuperK    &   & UNO \\ 
\hline
      Cavity Excavation        &    27,640 & v &  168,000 \\ 
      Water piping and pumps   &       630 & v &    4,082 \\ 
      Water Purification Sys   &     1,850 & v &   11,988 \\ 
      Power Station            &       720 & v &    2,160 \\ 
      Crane                    &       760 & v &    2,280 \\ 
      Water Tank               &    18,400 & s &   92,480 \\ 
      PMT support structure    &     4,580 & s &   23,019 \\ 
      Counting Room            &       330 & s &      990 \\ 
      Computer Building        &     1,860 & s &    2,232 \\ 
      Main Building            &     3,000 & s &    3,600 \\ 
      20\inch~PMT (including cables)   &    34,670 & s &  173,664 \\ 
      Electronics              &     6,330 & s &    9,495 \\ 
      DAQ                      &     1,090 & s &    1,635 \\ 
      Air Conditioning         &       210 & s &      315 \\ 
      Veto instrumentation     &     3,000 & s &    9,000 \\ 
      8\inch~PMT (including cables)   &     2,262 & s &   17,881 \\ 
      \hline 
      Total                    &   102,070 &   &  522,822 \\
\hline\hline
    \end{tabular} 
    \caption{UNO Cost Estimates (in thousands of US dollars): 
This is  a quite complete list of
expense items for all aspect of detector construction. 
The ``v'' and ``s'' symbols note the volume-like and 
surface-like nature of cost scalings used when the costs are
extrapolated from the Super-Kamiokande (SuperK) to UNO.   
A conversion rate of 1\$ = 100 yen is used.} 
    \label{tab:cost-estimates} 
\end{table} 

For the major cost items of the detector, I use more reliable quotes
and estimates for unit costs from vendors and experts. As shown in
Table~\ref{tab:estimated-unit-costs}, the unit cost for excavation is  
taken from the 
estimates by D. L. Petersen, and the unit costs for PMTs are taken
from Hamamatsu corporation. I take liberty of reducing the per
channel electronic cost by a factor of 5
and the water tank construction cost per unit surface by a factor of 2 
 from the Super-Kamiokande  cost.  
The cost of these five items comprise the majority of the UNO
detector cost.

\begin{table}[htbp] 
  \begin{tabular}{lll}
\hline\hline 
   Item & Unit Cost & Source \\
\hline 
    Excavation  & \$260/m$^3$   & L. Petersen \\ 
    20\inch~PMTs  & \$3,100       & Hamamatsu \\ 
    8\inch~PMTs  & \$1,200       & Hamamatsu \\ 
    Electronics & \$170/channel &  \\ 
    Water Tank  & \$2,076/m$^2$ &  \\ 
\hline\hline 
  \end{tabular} 
  \caption{Estimated Unit Costs: The excavation cost is assuming a
horizontal access tunnel and rock quality (Q value) of 100. The PMT
unit cost including cable cost is based on a 50k PMT order. It is \$2,850 if 
100k PMTs are ordered. A conversion rate of 1\$ = 100 yen is used.} 
  \label{tab:estimated-unit-costs} 
\end{table} 

\subsection{Cost Reduction Possibilities}

I believe that there is quite a bit of room to reduce the total 
cost for UNO. First of
all we can reduce the excavation cost by 
finding an existing cavity or underground facility. Or we could simply 
get a better quote. It is said that the actual excavation cost charged to the 
mining company in Kamioka, Japan is only \$50/m$^3$. 
Second, we could
use Mineguard type of liner rather than a stainless-steel container
for the container tank structure. Third, we should 
 optimize the PMT size and granuality for our physics goals and look
into a way of  developing 
new cheaper photo-detectors. For example, we could use two 8\inch~PMTs 
instead of a 20\inch~PMT, which may result in overall better
performance and cheaper cost. We could also look for other vendors
(from Russia?) than Hamamatsu. 

\subsection{Site Selection}

Obviously, the issue of site selection is an important issue for the
success of a project like UNO with this magnitude. 
In principle, a site near the Equator is desired if we are 
aiming for maximum information from UNO for the solar neutrino physics.
However, the consensus among solar neutrino theorists 
indicate that it is not crucial unless the MSW small mixing angle
solution looks very promising.  Thus, I concentrate my attention to
possible sites in the United States. 

An extensive  site search has been conducted by R.R. Sharp, Jr. and 
A. Mann during the early 1980s for a site for potential National Underground
Science Facility (NUSF). They surveyed mostly in the western states
for existing tunnels, inactive mines, active mines and new 
locations. They found a couple of sites in Nevada particularly 
attractive.\cite{Mann} Although the size and depth requirements 
of the potential sites they searched for are quite different from the   
UNO requirements, the information gathered is found to be quite useful.  

Another site search was conducted by W. Kropp {\it et. al.} 
concentrating on the site at 
San Jocinto, California. \cite{Kropp}
The results from this search also 
provides useful information for our current search.

The most attractive site so far presented to us is the WIPP (Waste
Isolation Pilot Project) site at Carlsbad, New Mexico. 
The facility  which is managed by DOE 
provides a hard salt rock geology, an already existing 
laboratory  facility and tunneling machinery. With strong government interest 
to utilize this facility for scientific research, the site is indeed
ideal for UNO or for 
any other large scale next generation underground detectors.

\section{Conclusions}

Feasibility of a 
next generation very large underground water Cherenkov detector, 
UNO, is considered. The detector configuration of UNO
is a linear compartmentalized detector (60mx60mx60mx3) with
a middle section with a 40\% photocathode coverage and two wing
sections with 10\% photocathode coverage. The middle section is  dedicated
for  low energy solar neutrino and supernova neutrino studies and for 
detecting 6 MeV prompt $\gamma$'s from $p \rightarrow \nu K^+$ 
decays.
 
The detector  
which has a total volume of 650 kton and a corresponding fiducial
volume of 445 kton (20 times larger than the fiducial volume 
of the Super-Kamiokande detector) can be built today at about
\$500M of construction cost. There are no known serious technical challenges
in building such a detector. Although rigorous 
R\&D is desired for cost reduction and improvement in detector
performance, there are no critical path R\&D items. 
If funding is available, we should be able to build this detector
within next ten years.

Such a detector, if built, will provide us with a bonanza of exciting
physics programs: Nucleon decay searches, precision  solar and 
atmospheric neutrino studies, and supernova neutrino observations. The 
detector can also serve as a far detector for a possible future 
neutrino factory. It will also compliment any other major accelerator 
based initiatives such as NLC or neutrino factories in terms of
providing diverse but crucial physics programs to 
the US HEP community. 

In my opinion, physics program of searches for nucleon 
decays undoubtedly belongs to ``MUST-DO'' physics category and it
should be vigorously pursued. I also believe if such a large
scale detector is built it should be a multi-purpose general detector
with a robust multi-decay mode search capability for nucleon decay
searches. It should not be specialized single purpose detector with
specific decay mode search capability.

\section{Acknowledgements}

This work is supported by the funding from the U.S. Department of
Energy under the contract No. DEFG0292ER40697. 
I wish to express my gratitude to 
Kenzo Nakamura who provided vital information on the Super-Kamiokande 
construction costs, D. Lee Petersen who provided information on
the technical details concerning underground excavation, Al Mann 
who provided information on the underground site searches. I also
like  to thank  John Bahcall, Bill Marciano, Jogesh Pati and Frank Wilczek 
for their theoretical inputs and encouragements, and Maurice
Goldhaber and other workshop organizing committee members for their
tireless work in  making the workshop successful and support for this work.


\begin{thebibliography}{99}
\bibitem{Pati}
K.S. Babu, J. C. Pati and F. Wilczek,
hep-ph/9812538; also see the articles in this proceedings
by Pati, by Wilczek and by Babu.
\end{thebibliography}
\end{document}